\documentclass[a4paper,11pt,reqno]{amsart}
\pdfoutput=1

\usepackage[]{amsmath}
\usepackage{amsfonts}
\usepackage{amssymb}
\usepackage{xcolor}
\usepackage{mathrsfs}
\usepackage{cancel}

\def\oemO{\overline{\emph{1}} }
\def\oemT{\overline{\emph{2}} }

\def\emO{{\emph{1}}}
\def\emT{\emph{2}}

\def\ob{\overline{1}}

\def\oltwo{\overline{2}}

\usepackage{tikz}
\usepackage[margin=3cm]{geometry} 
\usepackage{hyperref}

\theoremstyle{definition}

\renewcommand{\i}{{\mathrm{i}}}
\def\opsi{ \overline{\psi} }

\def\oPsi{ \overline{\Psi}{} }
\def\opsi{ \overline{\psi}{} }

\def\ttm{ \texttt{m}}
\def\ttw{ \texttt{w}}

\def\one{\hbox{{1}\kern-.25em\hbox{l}}}

\def\nn{\nonumber}

\def\oI{\overline{I}{}}

\def\oJ{\overline{J}{}}
\def\oLambda{\overline{\Lambda}{} }
\def\olambda{\overline{\lambda}{} }

\def\ob{\overline{b}{} }

\def\oK{\overline{K}{} }
\def\oB{\overline{B}{} }

\def\oM{\overline{M}{} }

\def\oPsi{\overline{\Psi}{} }

\def\btau{\mbox{\boldmath{$\tau$}}}
\newcommand{\Str}{\mathrm{Str}}
\newcommand{\Tr}{\mathrm{Tr}}

\newcommand{\cdt}{\!\cdot\!}

\def\DSlSl{D\hskip-7pt\slash\hskip-3pt \slash}

\def\PtSl{\partial\hskip-5pt\slash}

\definecolor{vertclair}{RGB}{0,153 0}

\definecolor{gris}{RGB}{153,153,153}

\title[FirstOrder ${} YMCT$] 
 {A first order nonabelian antisymmetric tensor model with hidden higher symmetry}
\author[FirstOrder ${} YMCT$]{Peter D. Jarvis${}^1$ and Jean Thierry-Mieg${}^2$}
\address{${}^1$School of Natural Sciences (Mathematics and Physics), University of Tasmania, PO Bag 37, Hobart, Tasmania 7001, Australia
(Alexander von Humboldt Fellow); 
\mbox{${}^2$NCBI}, National Library of Medicine, National Institute of Health, 8600 Rockville Pike, Bethesda MD20894, U.S.A.
}
\email{${}^1$peter.jarvis@utas.edu.au, ${}^2$mieg@ncbi.nlm.nih.gov}

\date{\today }

\begin{document}

\begin{abstract}\mbox{}\\
We present a nonabelian complex self-dual antisymmetric tensor model, with gauged internal symmetry, comprising a complex multiplet of massless self-dual antisymmetric tensor fields and its conjugate (interacting with massless Yang-Mills gauge fields). We build the model at the classical level, via a first order formulation, assigning local fields to a component expansion
over the Dirac algebra and a basis of internal matrices, with Lorentz and gauge invariance ensured by tracing, and applying an appropriate bilinear form. The full expansion requires auxiliary place-holder fields, which are not 
potentials, but become curvatures
(field strengths) after elimination via their algebraic equations of motion. The component expansion over the Dirac algebra is closed under multiplication by additional internal matrices, which are not gauge group generators, but mix internal components, and link auxiliary curvatures with potentials. The model is shown to be invariant under such transformations, and thus possesses a type of hidden internal symmetry. We analyze the specific case of gauge group $SU(2)\!\times \!U(1)$, and
as higher symmetry candidates, we consider both the Lie group $SU(3)$ and the Lie superalgebra $SU(2/1)$ (with the additional matrices acting via Lie brackets, or graded Lie brackets, respectively).

\end{abstract}
\maketitle
\vfill
\tableofcontents
\vfill

%\footnotetext{${}^*$ Alexander von Humboldt Fellow}
%\tableofcontents
\pagebreak
\section{Introduction and main results} 
\label{sec:Intro}
\mbox{}
In recent work we have re-visited the systematics \cite{jarvis2025antisymmetric} and conformal invariance properties \cite{thierry2025conformal} of antisymmetric tensor fields, and
the proposition that they may have a role to play as elements of a generalized standard model, together with the established fundamental fields and known particle spectrum. In particular, in an analysis of all couplings of antisymmetric
tensor fields in four dimensions, the known models, Ogievetsky-Polubarinov and Kalb-Ramond \cite{ogievetsky1967notoph,KalbRamond1974} and Avdeev-Chizhov \cite{AvdeevChizhov1994,Chizhov2011}, were confirmed, while also identifying a variant  Avdeev-Chizhov model with antisymmetric couplings, involving a pair of self-dual and anti self-dual tensors \cite{AvdeevChizhov1994,Chizhov2011}. These (free field) models were reconstructed in first order Duffin-Kemmer-Petiau 10 component formalism.   

In a remarkable early paper exploring ``higher spin'' fields, Green \cite{green1953first} pursued a similar investigation, also in a first order approach, but using a reducible, 16-component realization of the Kemmer matrices, via a two-sided action of Dirac matrices on elements of the Dirac algebra. This resulted in a same-mass Proca (10 component, vector spin-1) and Klein-Gordon (5 component, scalar  spin-0) system (together with a redundant scalar field). A generalization exploiting the Kemmer enveloping algebra itself also yielded higher-spin antisymmetric tensor fields with a nondegenerate mass spectrum.

In this work we extend the construction to include local gauge symmetry. This leads to a first order system describing massless Yang-Mills
fields, in interaction with a set of massless self-dual and conjugate anti self-dual antisymmetric tensor fields.
Local fields are assigned to a component expansion
over the Dirac algebra and a basis of internal matrices, with Lorentz and gauge invariance ensured by tracing and applying an appropriate bilinear form. The full expansion requires auxiliary place-holder fields, which are not
potentials, but
become curvatures (field strengths) after elimination via
their algebraic equations of motion. The component expansion over the internal and Dirac algebras
is closed under multiplication by internal matrices which are not gauge
group generators, but mix internal components, and link auxiliary curvatures with 
potentials. The model is shown to be invariant under such transformations, and thus possesses a type of hidden higher internal symmetry. 
We analyze the specific case of gauged $SU(2)\times U(1)$, and
as higher symmetry candidates, we consider both the Lie group $SU(3)$ and the Lie superalgebra $SU(2/1)$ (with the additional matrices acting via Lie brackets, or graded Lie brackets, respectively).

In section \ref{sec:YMCT} below, the details of the field component enumeration over the Dirac and internal matrix algebras are developed, leading to the presentation of the first order Yang-Mills complex antisymmetric tensor Lagrangian both in trace form and its expansion in terms of local fields (Yang-Mills potentials and complex self-dual antisymmetric tensor fields). The following section \ref{sec:Superalgebra} discusses the action of additional matrices which complement the gauge algebra by mixing field components, in particular when these matrices themselves close under the Lie or graded Lie brackets of a hidden higher symmetry. Detailed derivations of results reported in
sections \ref{sec:YMCT} and \ref{sec:Superalgebra} for the particular case of gauged $SU(2)\times U(1)$\,
%$G \times U(1) = SU(2)\times U(1)$\,
are relegated to appendix \ref{sec:Derivations}. 
We conclude in section \ref{sec:Discussion} with further comments on the model, with reference to the possibility that the higher symmetry algebra is either $SU(3)$, or $SU(2/1)$ in relation to recent work on enlarging the system to include coupling to fermions (in multiplets of $SU(2/1)$, as evidenced by the spectrum of observed quarks and leptons), as well as scalar fields and symmetry breaking. 
Further details of notation and useful reference data are provided in the remaining appendices. 
 
\section{Complex self-dual antisymmetric tensor Lagrangian}
\label{sec:YMCT}
%\noindent

Our purpose in this paper is to extend the construction method used in \cite{green1953first}, for building 
first order Lagrangians via trace techniques on fields over the Dirac algebra, to include local gauge
symmetry, and specifically, for complex self-dual antisymmetric tensor fields with
internal symmetry, and developing our earlier papers \cite{jarvis2025antisymmetric,thierry2025conformal}.
Before presenting our specific model below, we here provide a more general motivation for the required generalization, of combining fields in a matrix construction, over both Dirac and internal algebras. 

A first step is to note that the potential, or connection form $\sum W_\mu dx^\mu$ as an exterior differential, 
has a natural counterpart $ \sum W_\mu \gamma^\mu$ over a corresponding Dirac or Clifford algebra (which itself can be considered as a Hopf algebra deformation or ``quantization'' of the exterior algebra). In the case of local gauge symmetry,
$W$ is simultaneously an object over the Lie algebra of the local gauge group $G$\,, 
$W = \sum_{i=1}^N W^i \Lambda_i = \sum_{\mu=0}^3 W^\mu \gamma_\mu =\sum_{i, \mu} W^{i\mu} \Lambda_i \gamma_\mu$ in some basis. In combination with additional fields $T$ (chosen to be antisymmetric tensors - see below), carrying some $M$-dimensional representation of the internal gauge group, we
expand $T$  as  $T= \sum_{a=1}^M T^a \Lambda_a$,
where the $\Lambda_a$ are a list of standard basis vectors\,.
%
%$(\Lambda_i \cdt T) = \sum_{a,b} \Lambda_b (\Lambda_i)^b{}_a T^a$\,.

In this formulation, in order for the correct nonabelian curvature and covariant derivative definitions to emerge at the level of local fields, it must be possible for the internal ingredients to admit a multiplication
(for example, we shall see that the product $WW$ is tempered by the Dirac traces occurring, to encode the required matrix
commutator $[W_\mu,W_\nu]$ building the field strength). In the case of the local fields $T$, in order to arrange the required action 
$T^a \rightarrow (\Lambda_i)^a{}_b T^b$, it is reasonable to promote the vector basis
$\Lambda_a$ to a set of $M$ column matrices, so that we have the full expansion
\footnote{In the following, we write such expansions as $\Phi = \sum W^i \Lambda_i  + \sum T^a\Lambda_a$ with 
$\Lambda_i$, $\Lambda_a$ interpreted (by abuse of notation) in the sense of block matrices. Compare 
(\ref{eq:2Mplus1Scheme}), (\ref{eq:Mplus1SuperScheme}) and also (\ref{eq:ComplexTexpB}), (\ref{eq:Complextexp}) below. }
\begin{align}
\label{eq:Mplus1SchemeSimple}
%\Phi = \sum W^i \Lambda_i  + \sum T^a\Lambda_a \cong
\Phi = &\,
\sum W^{i}\left(\begin{array}{c|c} \Lambda_i & 0 \\ \hline 0 & 0
\end{array} \right)
\!+\! \sum T^a\left(\begin{array}{c|c} 0 & \Lambda_a \\ \hline 0 & 0
\end{array} \right)\,.
\end{align}
A special case is when the additional fields form a reducible representation, transforming as a complex conjugate pair, $B$ and $\oB$. 
The fields can as before be assembled, now as a column of length $2M$, in a complex basis $\Lambda_I ,  \oLambda_{\overline{I} }\,$, with $I  = \emO,\emT,\cdots, M\,$\,; $\oI = \oemO, \oemT, \cdots, \overline{M}$ equivalent to  
the $\Lambda_a$\,. Promoting these objects to $(2M\!+\!1)\times(2M\!+\!1)$ matrices as before, the full field array must now be expanded over both the gauge group generators and their contragredients (negative transpose), in block form
\begin{align}
\label{eq:2Mplus1Scheme}
\Phi \cong \sum W^{i} \left(\begin{array}{cc|c}  \Lambda_i & 0 &  0 \\
 0 & (-\Lambda_i)^\top & 0 \\
 \hline %\\%[.1cm] % &&\\ 
 0 & 0 & 0
\end{array} \right)\, +
B^I  \left(\begin{array}{cc|c} 0 & 0 & \Lambda_I \\
 0 & 0 & 0 \\
 \hline %\\%[.1cm] % &&\\ 
 0 & 0 & 0
\end{array} \right)\,+
\oB^{\overline{I}} \left(\begin{array}{cc|c} 0& 0 & 0 \\
 0 & 0 &\oLambda_{\overline{I} }\\
 \hline %\\%[.1cm] % &&\\ 
 0 & 0 & 0
\end{array} \right)\,,
%\qquad \Omega \cong \sum \left(\begin{array}{c|c} w^{i}\Lambda_i & k^a \Lambda_a \\\hline 0 & 0
%\end{array} \right)\,.
\end{align}

A more compact arrangement can be exploited if the 
matrices of the gauge generators act both as left- and, with sign reversed, right-multiplications (negative transpose) on the
matrices labelling the additional (tensor) field,
thus avoiding the doubling, and recovering a variant of the original $(M\!+\!1)\times(M\!+\!1)$ scheme
wherein the $\Lambda_I$ are column matrices and the $\oLambda_{\oI}$ become row matrices:
\begin{align}
\label{eq:Mplus1SuperScheme}
%\Phi \cong &\, \sum \left(\begin{array}{c|c}   W^{i}\Lambda_i + U \unit_M & H^I \Lambda_I 
\Phi \cong &\, 
\sum W^{i}\left(\begin{array}{c|c}   \Lambda_i  & 0
\\[.2cm] \hline
 \\[-.2cm] 
 0 & 0 
\end{array} \right)+
\sum B^I \left(\begin{array}{c|c}   0 & \Lambda_I 
\\[.2cm] \hline
 \\[-.2cm] 
0 & 0 
\end{array} \right)+
\sum \oB^{\oI}  \left(\begin{array}{c|c}   0 & 0 
\\[.2cm] \hline
 \\[-.2cm] 
\oLambda_{\oI} & 0 
\end{array} \right)\,.
\end{align}
Given that $\Lambda_i \oLambda_{\oJ} = 0=\Lambda_J \Lambda_i$, it is clear that the left- and right-multiplications
realizing the representation $B\rightarrow \Lambda_i B$\,,  $\oB\rightarrow -(\Lambda_i)^\top \oB= -\oB\Lambda_i  $\,, can in fact be regarded as 
arising from parts of corresponding matrix commutators, $[W,B]$\, or $[W,\oB]$\,, %as expected for covariant derivatives, namely
\begin{align}
\label{eq:RepsToComms}
[\Lambda_i, \Lambda_a] =&\, C_{ia}{}^b \Lambda_b\,, \quad \mbox{with}\quad C_{ia}{}^b := (\Lambda_i)^b{}_a\,,
\nonumber \\
\mbox{or} \quad
[\Lambda_i, \Lambda_I] =&\, C_{i I}{}^J \Lambda_J \,, \quad \mbox{and}\quad 
[\Lambda_i, \Lambda_{\oI}] = C_{i {\oI}}{}^{\oJ} \Lambda_{\oJ}\,.
\end{align}
This observation is suggestive of an extended algebraic picture,
incorporating both the generators $\Lambda_i$ of the gauge group as well as the internal matrices $\Lambda_a \sim \Lambda_I, \oLambda_{\oI}$\,. Indeed, in (\ref{eq:Mplus1SchemeSimple}) or
(\ref{eq:2Mplus1Scheme}), the full algebra is that of a semidirect product of the gauge group with an $M$- or  $2M$-dimensional abelian Lie algebra (or ``anti-abelian'' anticommuting Lie superalgebra) associated with the additional internal generators. 
In (\ref{eq:Mplus1SuperScheme}) however, it is evident that the internal generators may multiply gauge generators, with the effect of permuting internal assignments, or possibly multiply together, to return the gauge generators, provided that we add to (\ref{eq:Mplus1SuperScheme}) one further diagonal matrix generator, with nonzero entry in the lower-most block, together with its accompanying vector potential. We interpret this as an expansion of the gauge group to $G\times U(1)$\,.

It is this picture which we wish to develop here for the Yang-Mills complex self-dual antisymmetric tensor system, using the foregoing
internal arrangement, and with the local fields taking values in the Dirac algebra. The 
full ``Clifford field'' expansion is
\begin{align}
\label{eq:CliffSplits}
\Phi + \Omega := &\, (W + T) + (\omega + t) \equiv (W+\omega) + (T + t) =:\Phi_{YM} + \Phi_{CT}\,,
\end{align}
here displaying different separations either into ``potential'' (gauge and antisymmetric tensor) fields ($\Phi \sim W,T$)\, plus 
``auxiliary'' (curvature) fields ($\Omega \sim \omega, t$)\,, or into ``Yang-Mills'' ($\Phi_{YM} \sim W,\omega$)\,
plus ``complex tensor'' ($\Phi_{CT} \sim T, t$)\, component fields, respectively. Explicitly we have
\begin{align}
\label{eq:FullCliff}
\Phi_{YM} = W + \omega =
&\,\left.\sum\right._{m} W^{m \mu}\Lambda_m \gamma_\mu + \left.\sum\right._{m} \omega^{m \mu\nu}
\Lambda_m \tfrac 12 \sigma_{\mu\nu}\,,\\
\Phi_{CT} = T + t =
&\,\left.\sum\right._{a} T^{a \mu\nu}\Lambda_a \tfrac12 \sigma_{\mu\nu}+ \left.\sum\right._{a} t^{a\mu} \Lambda_a\gamma_\mu\,.\nonumber
\end{align}
\noindent 
Here we note a refinement of  (\ref{eq:CliffSplits}) above, which exploits the freedom in allocating antisymmetric tensor fields in the complex basis. Consider the expansion:
\begin{align}
\label{eq:ComplexTexpB}
\sum T^{a \mu\nu}\tfrac12 \sigma_{\mu\nu}\Lambda_a 
=&\, \sum B^{I\mu\nu} \Lambda_I \tfrac12 \sigma_{\mu\nu}\tfrac12(1\!+\!\gamma_5)
+ \sum \oB^{\oI \mu\nu}\oLambda_{\oI} \tfrac12 \sigma_{\mu\nu}\tfrac12(1\!-\!\gamma_5)\,.
\end{align}
The identity \smash{$\tfrac 12 \varepsilon_{\mu\nu}{}^{\rho\sigma} \sigma_{\rho\sigma} = \i \sigma
_{\mu\nu} \gamma_5$} (appendix \ref{sec:Derivations}) ensures that  
the complex internal multiplets  $ B^I$, $ \oB^{\oI}$ 
are assigned to distinct Lorentz representations, namely self-dual
and anti self-dual, respectively\footnote{
%The factor $\times \tfrac12$ is chosen for later correct normalization of the action. 
We define a self-dual tensor $B$ (in Minkowski space) 
 by the condition ${}^*B_{\mu\nu} := +\i B_{\mu\nu}$, where ${}^*B_{\mu\nu} := \tfrac 12 \varepsilon_{\mu\nu}{}^{\rho\sigma}B_{\rho\sigma}$. Then  $\tfrac 12 B \cdot \sigma = \tfrac 14 (B -\i {}^*B \cdot \sigma)=\tfrac 12 B \cdot \sigma \tfrac 12(1+\gamma_5)$\,. }. We use this allocation of component fields from now on, but continue to write
simply \smash{$\left.\sum\right._{a} T^{a \mu\nu}\tfrac12 \sigma_{\mu\nu}\Lambda_a$}\, to mean (\ref{eq:ComplexTexpB}), and similarly 
\smash{$\left.\sum\right._{a} t^{a \mu} \gamma_{\mu}\Lambda_a$} to mean
\begin{align}
\label{eq:Complextexp}
\left.\sum\right._a t^{a \mu}\Lambda_a \gamma_{\mu}
=&\, \left.\sum\right._I b^{I\mu} \Lambda_I \gamma_\mu
+ \left.\sum\right._{\oI} \ob^{\oI \mu}\oLambda_{\oI} \gamma_\mu\,.
\end{align}
in the complex basis.

In the construction of the local Lagrangian,  we use the notation
%$\langle \varphi_1\cdt \varphi_2 \rangle$ on ``Clifford field'' components
$\langle (\,\cdt\,)\cdt (\,\cdt\,) \rangle$ on ``Clifford fields''  
to denote 
both the trace over Dirac matrices, and the imposition of an appropriate even invariant bilinear form. Thus
``$\langle \varphi_1\cdt \varphi_2 \rangle$'' on components will denote
\[
\langle (\varphi_1(x) \Lambda_1 \Gamma_1)\cdt (\varphi_2(x) \Lambda_2\Gamma_2) \rangle
:=  
\varphi_1(x)\varphi_2(x) \, \zeta_{\Lambda_1 \Lambda_2}\, \Tr(\Gamma_1\Gamma_2)\,,
\]
where $\varphi_1(x)\varphi_2(x)\Tr(\Gamma_1\Gamma_2)$ determines the Lorentz structure, and  
with internal invariant bilinear form $\langle \Lambda_1\cdt \Lambda_2 \rangle=\zeta_{\Lambda_1 \Lambda_2}$ having $N\!\times\!N + 1\!\times\!1$ block form over the gauge generators, $\zeta_{mn} = \zeta_{nm}$, $m,n = 1,2,\cdots, N\!+\!1$, and also $2M\! \times\! 2M$ block form $\zeta_{ab}$ over the additional internal generators, or in the complex basis with nonzero components $\zeta_{I \oJ}$\, and  $\zeta_{\oJ I}$, $I, \oJ =  \emO, \oemO, \emT, \oemT, \cdots, M, \oM$\,.

In the rest of this work we fix $G=SU(2)$, with extended gauge group $SU(2)\times U(1)$, that is\footnote{That is, the internal expansion is 
over $3\times 3$ matrices, $M+1=3$, and the extended gauge group has dimension $N+1=4$\,.} $N=3$\,, and $M + \oM=2 + \overline{2}$, with the nonzero $2\times \oltwo$ and $\oltwo \times 2$ invariants proportional to the standard antisymmetric matrix. 
Potential ``hidden higher symmetries'' of the action will be defined, and the invariance properties established, in section \ref{sec:Superalgebra} below where we 
consider as minimal candidates the Lie algebra $SU(3)$ and Lie superalgebra $SU(2/1)$\,.
In the concluding section \ref{sec:Discussion} further technical aspects and the physical relevance of the model will be discussed. 

We propose 
 \begin{align}
 \label{eq:CliffordFieldL}
\qquad {\mathscr L}_{YMCT} = &\,
\langle(\i \DSlSl \Phi)\cdt \Omega \rangle + \langle \Omega\cdt \Omega\rangle \nonumber\\
=&\, 
\langle \big(\i\PtSl + W-T )(W +T)\big)\cdt (\omega+ t)\rangle + \langle(\omega+t)\cdt(\omega+t)\rangle\,,
 \end{align}
 where $\i \DSlSl \Phi = (\i\PtSl + W-T)\Phi := \Psi $ is the extended covariant derivative. Here (anticipating which terms will provide nonzero Dirac traces) the internal structure is
 \begin{align}\
 \label{eq:ZetaChoice}
 \langle \Psi\cdt \Omega \rangle =&\,
 \left.\sum\right._{k,\ell} \Tr(\Psi^k \omega^\ell) \zeta_{k\ell} + 
 \left.\sum\right._{I,\oJ} \Tr(\Psi^I \ob^{\oJ}) \zeta_{I \oJ} + \left.\sum\right._{\oI, J} \Tr(\oPsi^{\oI} b^{J})\zeta_{\oI J} \,,
\\
\mbox{but}\qquad 
 \langle \Omega\cdt \Omega \rangle =&\,  \langle \omega\cdt \omega \rangle+ \langle t\cdt t \rangle=
 \left.\sum\right._{k,\ell} \Tr(\omega^k \omega^\ell) \zeta_{k\ell}
 + 2 \left.\sum\right._{I,\oJ} \Tr(b^I \ob^{\oJ}) \zeta_{I \oJ} \,,\nonumber
\end{align}
for the homogeneous bilinear.

We proceed with the presentation for the $SU(3)$ higher symmetry, with the $SU(2/1)$ case to be treated separately below.  For definiteness the bilinear form $\zeta_{mn}$, $\zeta_{ab}$ will be denoted $\eta_{mn},\eta_{ab} $\,, and $\kappa_{mn},\kappa_{ab}$\, in the Lie algebra and Lie superalgebra cases, respectively (see below). 
                               
Evaluating the nonvanishing Dirac trace and internal bilinear invariant combinations leads in the $SU(3)$ case to
(appendix \ref{sec:Derivations})
\begin{align*}
\qquad 
 {\mathscr L}_{YM} :=&\, \langle \tfrac 12(\i\PtSl W + WW) \cdt \omega\rangle + \langle\omega\cdt\omega\rangle
 = \big(F^{m\mu\nu}(W)\omega^n_{\mu\nu} + 2 \omega^{m \mu \nu} \omega^{n}_{\mu \nu}\big)\eta_{mn}\,.
   \end{align*} 
For the self-dual antisymmetric tensor Lagrangian we find
\begin{align*}
{\mathscr L}_{CT} := &\,\tfrac12 \langle (\i \PtSl T + [W,T])\cdt t\rangle + \langle t \cdt t \rangle
= -4 (D_\mu T)^{a\mu\nu} t^b{}_\nu \eta_{ab}+ 4 t{}^{a\mu}t^{b}{}_{ \mu}\eta_{ab}
%\\
%\Rightarrow &\,{\mathscr L}_{CT} =-\tfrac 12 (D_\rho T')^{a\rho \nu} (D_\sigma T)^{b\sigma}{}_{ \nu} \eta_{ab}\,.
%\qquad\qquad
\end{align*}
The local Lagrangian ${\mathscr L}^1_{YMCT}={\mathscr L}^1_{YM} + {\mathscr L}^1_{CT}$ at first order is thus
\begin{align}
\label{eq:finalL_ymac_first}
{\mathscr L}^1_{YMCT}=&\,
F^{m\mu\nu}(W)\omega^n_{\mu\nu}\eta_{mn} + 2 \omega^{m \mu \nu} \omega^{n}_{\mu \nu} \eta_{mn}
-4 (D_\mu T)^{a\mu\nu} t^b{}_\nu \eta_{ab}+ 4 t{}^{a\mu}t^{b}{}_{ \mu}\eta_{ab}\,. 
\end{align}
At second order (completing the square and removing a Gaussian from the generating functional, or eliminating 
fields using the algebraic constraints), the auxiliaries $\omega^k_{\mu\nu}$ and $t^a{}_\mu$ are removed in favour of field strengths, viz. $\omega^k{}_{\mu\nu} = -\tfrac 14 F(W)^k{}_{\mu\nu}$\,,  and $t^a{}_\mu =\tfrac 12 DT^a{}_\mu$\,.
The result is
\begin{align}
\label{eq:finalLymacSecond}
{\mathscr L}^2_{YMCT}=&\,-\tfrac 18(F(W)^{m \rho \sigma} F(W)^{n}{}_{ \rho \sigma}\eta_{mn} 
%- s D\!B^{I \rho} D\!\oB^{\oJ}{}_\rho \zeta_{I\oJ}\,.
- (D\cdt T)^{a \rho}(D\cdt T)^{b}{}_{ \rho}\eta_{ab}\\
\equiv &\, -\tfrac 18(F(W)^{m \rho \sigma} F(W)^{n}{}_{ \rho \sigma}\eta_{mn} -2D\cdt B^{I\rho} D\cdt\oB^{\oJ}{}_\rho \eta_{I\oJ}
\nonumber
\end{align}
In this form ${\mathscr L}^2_{YMCT}$, up to overall scaling, is identical to the nonabelian complex self-dual 
extension \cite{LemesRenanSorella1995II} (without tensor self-coupling) of the original Avdeev-Chizhov model of  \cite{AvdeevChizhov1994,Chizhov2011}.

In the case of  $SU(2/1)$ as candidate higher symmetry, we must use the complex basis, and note several
important differences in the structure of the bilinear form and also the structure constants 
(appendix \ref{app:NotationSU21SU3}, compare (\ref{eq:TripletLambdas}) - (\ref{eq:KillingOddLAComp})
with (\ref{eq:KillingEven}) - (\ref{eq:KappaRealFbyF})).
Eliminating the auxiliaries from 
\begin{align}
\label{eq:superL1complex}
{\mathscr L}^1_{YMCT}=  &\, {\mathscr L}^1_{YM}+ {\mathscr L}^1_{CT}
= F(W)^{m \rho \sigma} \omega^{n}{}_{ \rho \sigma}\kappa_{mn}+2 \omega^{m \rho \sigma} \omega^{n}{}_{ \rho \sigma}\kappa_{mn}\\
&\,-4 (D_\mu B)^{I\mu\nu} \ob^{\oJ}{}_\nu \kappa_{I\oJ}
-4 (D_\mu \oB)^{\oI\mu\nu} b^{J}{}_\nu \kappa_{\oI J} +8 b^{I\mu\nu} \ob^{\oJ}{}_\nu \kappa_{I\oJ} \nonumber
\end{align}
 leads to\footnote{In terms of the standard $2\times 2$ antisymmetric matrix, we have
 $\eta_{I\oJ}=\varepsilon_{I\oJ}=\kappa_{I\oJ}$ but $\eta_{\oI J}=
 -\varepsilon_{I\oJ}$\,, $\kappa_{\oI J}=+ \varepsilon_{\oI J}$ leading to 
 (\ref{eq:finalLymacSecond}) and (\ref{eq:YMCTsuper}), respectively (appendix \ref{app:NotationSU21SU3}).}
\begin{align}
\label{eq:YMCTsuper}
{\mathscr L}^2_{YMCT}
=&\, -\tfrac 18(F(W)^{m \rho \sigma} F(W)^{n}{}_{ \rho \sigma}\kappa_{mn} +2D\cdt B^{I\rho} D\cdt\oB^{\oJ}{}_\rho  \kappa_{I\oJ}\,.
%
%-4 (D_\mu B)^{I\mu\nu} \ob^{\oJ}{}_\nu \kappa_{I\oJ}
%-4 (D_\mu \oB)^{\oI\mu\nu} b^{J}{}_\nu \kappa_{\oI J} +8 b^{I\mu\nu} \ob^{\oJ}{}_\nu \kappa_{I\oJ}
\end{align}
 
\section{Higher symmetry}
\label{sec:Superalgebra}
\mbox{}\\[-.7cm]
\noindent
In this section we revisit the Clifford field expansions in order to elucidate
the role the set of additional internal matrices $\Lambda_a$ or $\Lambda_{I}, \oLambda_{\oI}$ labelling the tensor fields may play in realizing a possible``higher symmetry" for the model. As noted, the gauge generators 
(apart from the additional $U(1)$ generator)
admit a one-sided multiplication rule (see (\ref{eq:Mplus1SuperScheme})), which in keeping with the Lie bracket structure, we promote to a commutator action (now including the additional $U(1)$ generator):
\[
[\Lambda_c, \Lambda_m] = C_{cm}{}^a \Lambda_a\,.
\]
This is of course the negative of the action of the commutator of the gauge generators with the additional internal matrices, with $C_{cm}{}^a = - C_{mc}{}^a$\,. 

Note now that the bipartitions (\ref{eq:CliffSplits}) amongst the four Clifford fields (\ref{eq:FullCliff}) 
admit a third split,
\begin{align}
\label{eq:CliffSplitsNo3}
\Phi + \Omega := &\, (W + T) + (\omega + t) \equiv (W+\omega) + (T + t) =:\Phi_{YM} + \Phi_{CT}\,\nn \\
\equiv &\,  (W+t) + (\omega + T)=:\Phi_{\gamma} + \Phi_{\sigma}\,,
\end{align}
reflecting the organization of fields by Lorentz structure (vector and tensor). In this way we can 
define a (global) algebraic-level transformation rule  on component fields\footnote{In appendix \ref{sec:Derivations}\,, in order to match dimensions correctly, we introduce explicit rescaling of the potentials in $\Phi_{\gamma}$ by fixed 
mass parameters, namely $W\rightarrow \ttm_W W$, $T\rightarrow \ttm_T T$ (or $B\rightarrow \ttm_{B} B$, $\oB \rightarrow \ttm_{\oB} \oB$ ). }
by imposing an appropriate 
bracket on $\Phi_{\gamma}$ and $\Phi_{\sigma}$ separately, thus
$\delta_c(\Phi_{\gamma}):= [\Lambda_c, \Phi_\gamma]$ and $\delta_c(\Phi_{\sigma}):= [\Lambda_c, \Phi_\sigma]$\,, or
\begin{align}
\label{eq:DeltaCcliff} 
\delta_c(\Phi_{YM\!+\!CT}) =&\,
\delta_c(W) + \delta_c(\omega) +\delta_c(T) +\delta_c(t) \\
:= &\,[\Lambda_c,\Phi_{YM\!+\!CT}] =
%= \left.\sum\right._m [\Lambda_c, W]+ \left.\sum\right._m[\Lambda_c, W] +\left.\sum\right._a +\left.\sum\right._a + 
[\Lambda_c, W] + [\Lambda_c, \omega] + [\Lambda_c, T] + [\Lambda_c, t]\,. \nonumber
\end{align}
The new ingredient here is the introduction of an extended algebraic structure such that the brackets with $T$ and $t$
will close on gauge generators, namely
\begin{align}
\label{eq:NewBracket} 
[\Lambda_c,\Lambda_a] = &\, C_{ca}{}^m \Lambda_m\,,
\end{align}
assuming that the enlarged algebraic system is consistent\footnote{In the arrangement 
(\ref{eq:2Mplus1Scheme}), closure is guaranteed trivially by the 
%mutual commutation or anticommutation of the additional generators,
fact that the additional matrix generators are nilpotent,
with vanishing $C$ coefficients, consistent with the semi-direct product of the gauge algebra sector with an 
abelian or ``anti-abelian'' part.}, 
leading to
\begin{align}
\label{eq:DeltaCcpts}
\delta_c(W^{m\mu}) =&\, t^{a\mu}C_{ca}{}^m \,,\quad \delta_c(\omega^{m\mu\nu}) = T^{a\mu\nu}C_{ca}{}^m\,, \\
\delta_c(T^{a\mu\nu}) =&\, \omega^{m\mu\nu}C_{cm}{}^a \,,\quad \delta_c(t^{a\mu}) =  W^{m\mu}C_{cm}{}^a  \,. \nonumber
\end{align}
The coefficients $C_{ca}{}^m$ in the first line (c.f. (\ref{eq:RepsToComms})) express 
the fact that the extra internal generators close on gauge group generators multiplicatively,
or indeed under \emph{either} a commutator \emph{or} an anticommutator
\footnote{From (\ref{eq:Mplus1SuperScheme})\,, in the complex basis, for generators of $G$, the $C_{I{\oJ}}{}^k$\,, $k=1,2,\cdots, N$ are identical, for both Lie algebra and Lie superalgebra
cases. The structure relation  $C_{{\oJ}I}{}^k = \mp C_{I{\oJ}}{}^k$ comes about here simply because the nonvanishing matrix multiplication involved occurs with a different sign in the definition of the commutator as compared with the anticommutator. Appendix \ref{app:NotationSU21SU3} provides a detailed expos\'{e} of the similarity
between the respective $3\times 3$ matrix structures of the Lie algebra and Lie superalgebra.}.
Note also the equivalents of 
 (\ref{eq:DeltaCcpts}) in the complex basis:
  \begin{align}
\label{eq:DeltaCcliffComplex} 
 \delta_I(W^{m\mu}) =&\, \ob^{\oJ\mu}C_{I{\oJ}}{}^m \,,\quad \delta_I(\omega^{m\mu\nu}) = {\oB}^{{\oJ}\mu\nu}
 C_{I{\oJ}}{}^m\,, \\
\delta_I(B^{J\mu\nu}) =&\, \omega^{m\mu\nu}C_{Im}{}^J \,,\quad \delta_I(b^{J\mu}) =  W^{m\mu}C_{Im}{}^J  
\,,\nonumber\\ 
\label{eq:DeltaCcliffComplex2}
\delta_{\oI}(W^{m\mu}) =&\, b^{J\mu}C_{{\oI} J}{}^m \,,\quad \delta_{\oI}(\omega^{m\mu\nu}) = B^{ J \mu\nu}
C_{{\oI} J}{}^m\,, \\
\delta_{\oI}(\oB^{{\oJ}\mu\nu}) =&\, \omega^{m\mu\nu}C_{{\oI}m}{}^{\oJ} \,,\quad 
\delta_{\oI}(\ob^{{\oJ}\mu}) =  W^{m\mu}
C_{{\oI}m}{}^{\oJ}  \nonumber
\,.
\end{align}
 
The sought-after ``higher symmetry" of the model is thus characterized by the component field transformation rules
\footnote{As written these transformation rules are incomplete, in that \smash{$\delta_I(B^{J\mu\nu})$} and 
\smash{$\delta_{\oI}(\oB^{\oJ\mu\nu})$} as well as \smash{$\delta_I(\omega^{m\mu\nu})$} and \smash{$\delta_{\oI}(\omega^{m\mu\nu})$} should involve the appropriate
self-dual and anti-self dual projections of the antisymmetric tensor $\omega$, which have been omitted here for brevity.
In addition, it should be noted that the transformation rules between potentials and curvature components also require
formal mass scaling parameters in order for the formalism to be consistent dimensionally. The
re-scalings $W\rightarrow \ttm_W W$, $T\rightarrow \ttm_T T$ (or $B\rightarrow \ttm_{B} B$, $\oB \rightarrow \ttm_{\oB} \oB$ ) are adopted in appendix 
\ref{sec:Derivations}. 
}
(\ref{eq:DeltaCcpts}) or equivalently (\ref{eq:DeltaCcliffComplex})\,, (\ref{eq:DeltaCcliffComplex2})\,, with
new coefficients $C_{ca}{}^m$\,, (or $C_{I \oJ}{}^m\,, C_{\oI J}{}^m$) which we take to be scalars, that is structure constants\footnote{We do not entertain $q$-deformations or generalized Witt constructions \cite{hannabuss1998quantum} in this work.}, for the choice of Lie algebra 
or Lie superalgebra in which the gauged algebra $SU(2)\!\times \!U(1)$ is naturally embedded, namely
$SU(3)$, or  $SU(2/1)$, respectively. We thus have two candidate schemes based on the closure of bilinear brackets, either via commutators (Lie brackets) or both commutators and anticommutators (graded Lie brackets for the superalgebra), respectively. See appendix \ref{app:NotationSU21SU3} for explicit definitions, notation and a compendium of basic data regarding these algebras. 
  
For convenience we summarize our notation as follows.
We take generators $\Lambda_i$ of $SU(2)$ in a Cartesian basis\,, $i,j,k = 1,2,3$\,, 
%(and later $m,n,p = 0, i,j,k = 0,1,2,3$\,; even)
and enumerate $\Lambda_a$ as $a,b,c = 4,5,6,7$\,. Unless otherwise specified, these
$\{\Lambda_i, \Lambda_a\}$  are the standard $3\times 3$  octet
Gell-Mann matrices $\tfrac12\lambda_1,\tfrac12\lambda_2,\tfrac12\lambda_3,   \tfrac12\lambda_4,  \tfrac12\lambda_5,   \tfrac12\lambda_6, \tfrac12\lambda_7$\,. In equation (\ref{eq:FullCliff}) above, the index set for gauge generators is $N\!+\!1$-fold, namely we have extended $i,j,k = 1,2,3$ to $m,n,p,q =  1,2,3,8$ for $SU(3) $, and $m,n,p,q = 0,1,2,3$ for $SU(2/1) $\,.
The additional $U(1)$ generator $\Lambda_8=\tfrac 12 \lambda_8$ for $SU(3)$ entails the standard traceless Gell-Mann matrix $1/\sqrt3\,\mathrm{diag}(1,1,-2)$, and $\Lambda_0$ for $SU(2/1)$ is the additional  super-traceless matrix $\Lambda_0=\lambda_0 =\mathrm{diag}(1,1,+2)$\,. See appendix \ref{sec:SprTensor} for tensor notation for superalgebras. We refer to the off-diagonal matrices $\Lambda_a$ as ``odd'', and to the diagonal matrices as ``even"\,. See appendix 
\ref{app:NotationSU21SU3} for definitions. In the complex basis, 
replacing $\Lambda_a\,, a=4,5,6,7$\, by $\Lambda_{\emO}, \Lambda_{\emT}, \oLambda_{\oemO}, \oLambda_{\oemT}$\,, for this special case of $G=SU(2)$\,,
we introduce two $SU(2)$ doublets transforming \emph{identically} but with opposite $U(1)$ charges (adjoint action of 
$\Lambda_8$ or $\Lambda_0$ respectively).

Here we demonstrate for the $SU(3)$ choice the invariance of the on shell action (\ref{eq:finalLymacSecond})\,,
\begin{align*}
%\label{eq:finalLymacSecond}
{\mathscr L}_{YMCT}=&\,-\tfrac 18(F(W)^{m \rho \sigma} F(W)^{n}{}_{ \rho \sigma}\eta_{mn} 
%- s D\!B^{I \rho} D\!\oB^{\oJ}{}_\rho \eta_{I\oJ}\,.
- (D\cdt T)^{a \rho}(D\cdt T)^{b}{}_{ \rho}\eta_{ab}%\\
%\equiv &\, -\tfrac 18(F(W)^{m \rho \sigma} F(W)^{n}{}_{ \rho \sigma}\eta_{mn} -2B^{I\mu\nu} \oB^{\oJ}{}_\nu \eta_{I\oJ}
%\nonumber
\end{align*}
where the above-derived transformation rules $\delta_c(\cdot)$ (or $\delta_{I}(\cdot)$, $\delta_{\oI}(\cdot)$ in the complex basis) between the  component potential fields and auxiliaries 
will be transcribed, by eliminating the latter in terms of their algebraic equations of motion, into transformation rules between potentials and field strengths. Using these, and dropping proportionality factors for this discussion, we have
\begin{align}
\label{eq:OddTransfsOnShell}
\delta_cW^m{}_\rho = &\, C_{ca}{}^m (D\cdt T)^a{}_\rho\,, \quad 
\delta_cT^a{}_{\rho\sigma} =   C_{cm}{}^a (F(W))^m{}_{\rho\sigma}\,.
\end{align}
From (\ref{eq:OddTransfsOnShell}) we have the expansion 
\begin{align*}
\delta_c F(W)^m{}_{\rho\sigma}F(W)^{n\rho\sigma}\eta_{mn} =&\, 2\big(\partial_\rho \delta_cW^m{}_\sigma + 
C_{k\ell}{}^m W^k{}_\rho\delta_cW^\ell{}_\sigma \big)F(W)^{n\rho\sigma}\eta_{mn}
\end{align*}
which gives
\begin{align*}
\delta_c F(W)^m{}_{\rho\sigma}=&\,  D_\rho \delta_cW^m{}_\sigma -D_\sigma \delta_cW^k{}_\rho\,,
\end{align*}
and
\begin{align}
\label{eq:PartialToDdiff}
\delta_c F(W)^m{}_{\rho\sigma}F(W)^{n\rho\sigma}\eta_{mn} =&\,2 C_{ca}{}^m (D_\rho (D\cdt T)^a{}_\sigma)F(W)^{n\rho\sigma}\eta_{mn}\,.
\end{align}
Similarly from (\ref{eq:OddTransfsOnShell}) we expand
\begin{align*}
\delta_c(D\cdt T)^a{}_\rho = &\, C_{cm}{}^a (D\cdt F)^m{}_\rho +
C_{me}{}^a C_{cf}{}^m (D\cdt T)^f{}_\sigma T^{e\sigma}{}_\rho\,,
\end{align*}
yielding
\begin{align}
\label{eq:DeltaDdotTDT}
\hskip-.5cm\delta_c(D\!\cdt \!T)^a{}_\rho(D\!\cdt \!T)^{b\rho}\eta_{ab} = &\, C_{cm}{}^a (D\!\cdt \!F)^m{}_\rho(D\!\cdt \!T)^{b\rho} \eta_{ab}\!+\!
C_{me}{}^a C_{cf}{}^m (D\!\cdt \! T)^f{}_\sigma T^{e\sigma}{}_\rho(D\!\cdt \! T)^{b\rho}\eta_{ab}\,.
\end{align}
At this point we invoke the dynamical Euler-Lagrange equations of motion for the potentials $W^{m\mu}$ and
$T^{a\mu\nu}$\,, which read respectively
\begin{align}
\label{eq:ELequationW}
D_\rho \omega_m{}^{\rho \mu}= &\, \tfrac 12 C_{mab} T^{a\mu\sigma}t^{b}{}_{\sigma}\,, \,\mbox{or}\,
 (D\cdt F)_{m}{}^{\mu} =  C_{mab} T^{a\mu\sigma}(D\cdt T)^{b}{}_{\sigma}\,,
 %:= (T D\cdt T)^\mu{}_m\,, 
\,\, \mbox{and}\\
\label{eq:ELequationT}
D_\rho t^a{}_\sigma = &\,D_\sigma t^a{}_\rho\,, \,\,\mbox{or}\, \, 
D_\rho (D \cdt T)^a{}_\sigma = D_\sigma (D \cdt T)^a{}_\rho\,. 
\end{align}
From the $T$ equation (\ref{eq:ELequationT}), it is evident that (\ref{eq:PartialToDdiff}) vanishes, or equivalently
(by partial integration $C_{cam}D\cdt F^{m\rho} D\cdt T^a{}_\rho=0$, so that the first term of (\ref{eq:DeltaDdotTDT}) also vanishes. The second, $CC$ term of  
(\ref{eq:DeltaDdotTDT}) can also be written as $(D\cdt F)(D\cdt T)$\,, by using the $W^{m\mu}$
equation of motion (\ref{eq:ELequationW}) to substitute $(T D\cdt T)$ for $(D\cdt F)$ (with the same index structure) which again 
leads to the vanishing $(D\cdt F)(D\cdt T)$ combination.  Analysis of the on shell $SU(2/1)$ invariance of the second order action (\ref{eq:YMCTsuper}) under $\delta_{I}(\cdot)$, $\delta_{\oI}(\cdot)$  
proceeds similarly. An equivalent demonstration of the invariance of the off shell first order action
is given in appendix \ref{sec:Derivations}\,. 

 %\pagebreak
\section{Summary and concluding remarks}
\label{sec:Discussion}
%\mbox{}
%
%\noindent
In this paper we have provided a Lagrangian model of nonabelian complex self-dual and anti-self dual antisymmetric tensors in four dimensions, in a first order formalism with fields expanded over both the Dirac 
algebra, and a matrix basis including both gauge group generators and additional internal matrices.

Remarkably, the method of dressing component fields by elements of the Dirac algebra (``Clifford fields'') in our construction permits the emergence of additional symmetry transformations which mix gauge and tensor fields, not with each other, but rather with their respective curvature fields.
The hidden ``higher symmetry" thus acts as a type of ``chaperone'' accompanying the gauge group.
In the text we have retained, as admissible candidates, either a larger Lie group, or (up to caveats to be discussed presently), an associated Lie superalgebra.  

The action in our present model has been shown to be invariant under such additional transformations for either choice. While the formalism extends naturally to $SU(n)\!\times\! U(1)$ and higher symmetries $SU(n\!+\!1)$ or $SU(n/1)$, we have examined in detail the simplest test case of gauged $SU(2)\!\times\! U(1)$, and higher symmetries $SU(3)$ or $SU(2/1)$. 
For example, we may conceive $SU(3)$ in the role of chaperone symmetry for an anomalous ``colour'' dynamics:
fermion triplets (quarks), with nonabelian massless \emph{dichromatic} $SU(2)\!\times\! U(1)$ gluons, in interaction with massless complex self dual tensors. As an extended example, an $SU(6)$ group might be imagined as chaperone to a grand unified $SU(5)\!\times\! U(1)$ gauge symmetry: in this case, the standard ``lepto-quark'' vector gauge bosons of $SU(5)$ would be accompanied also by ``lepto-quark'' tensor bosons\footnote{Compare 
\cite{dondi1980assignments}}.

We now turn to the case for a superalgebra in the role of a chaperone symmetry in the particle spectrum. Indeed, a motivation for the present work has been to contribute to a plausible foundation for the 
 almost 50 year-old suggestion that ``internal supersymmetry'' may play a role\footnote{S. Wouthuysen, private communication (c.f. \cite{Wouthuysen1978}).}, and in particular that the field assignments of the standard electroweak model appear to fall naturally into low-dimensional representations of the Lie superalgebra $SU(2/1)$, hinting at an anomalous role for such
 an internal supersymmetry, and flying in the face of long-held basic principles of quantum field theory. 
 Early works on this aspect \cite{fairlie1979higgs,ne1979irreducible,dondi1979supersymmetric} noted the correspondence between the spectrum of electroweak gauge bosons plus complex isodoublet scalar fields
 and the octet adjoint representation of $SU(2/1)$ (with the scalars' and vectors' spin difference being attributed
 to an artefact of a supersymmetry assignment of non Fermi-Bose type). At the same time it was also noticed that the observed spectrum of
 left-and right-handed lepton triplets and quark quartets conformed naturally with low-dimensional multiplets of $SU(2/1)$, with hypercharge as a symmetry generator, the spin difference between left- and right-handed fermions again 
purported to be a consequence of a graded supersymmetry. Subsequent works \cite{neeman1980geometrical,thierry1982exterior} have proposed a generalized differential geometric description (c.f. \cite{quillen1985superconnections} ) in which an inhomogeneous  exterior form includes scalar fields and also antisymmetric tensors as well as gauge potentials, while respecting the supersymmetry. A striking result in this framework 
\cite{thierry2020scalar,ThierryMieg2021newchi,ThierryMiegJarvis2021newchi} is that couplings to quarks and leptons not only reproduce the standard chiral anomaly cancellations at one loop, but also yield related chiral scalar and tensor 
anomaly cancellations. Together these can be identified as the vanishing of different components of the cubic $SU(2/1)$ Casimir tensor for the physical fermion spectrum, amounting to a vast generalization of standard anomaly cancellation to a super-BIM mechanism. The role of $SU(2/1)$ as a physically relevant symmetry structure has also been noted in connection with its finite-dimensional representations: not only do the standard leptons and quarks fall into irreducible multiplets of $SU(2/1)$ (either triplets or quartets), but there also exist indecomposable copy representations \cite{marcu1980tensor,Su1992,germoni1998indecomposable,jarvis2022indecomposable,thierry2022explicit}
which have been suggested \cite{coquereaux1991elementary} as a receptacle for repeated family generations (and potentially, their mixing parameters).

The above context pertains to the role of $SU(2/1)$ both as a kinematical symmetry, and in interactions. A detailed account of the latter (one loop) results is beyond the scope of this paper. However, recall that the decoration of component fields by elements of the Dirac algebra, leading to 
the Clifford field expansions (\ref{eq:CliffSplitsNo3}) in the present model, was instigated by replacing the exterior differential $W_\mu dx^\mu$  by the Dirac equivalent $W_\mu \gamma^\mu$\,. Thus this model (or possible extensions to it) is naturally adapted to the introduction of $SU(2/1)$ invariant interactions with physical fermions $\psi$  in irreducible (or indecomposable) representations, namely left-and right-handed spinors in supermultiplets graded by $\gamma_5$ chirality\footnote{In 
\cite{jarvis2025antisymmetric} the self-dual and anti-self dual antisymmetric tensors are characterized in the 10 component first order formalism as $\pm \i $ eigenstates of the pseudoscalar Kemmer matrix $\beta_5 := \beta_0 \beta_1 \beta_2 \beta_3$\,.}: 
the extended covariant derivative term in equation (\ref{eq:CliffordFieldL}) should simply be taken over 
as the term $\opsi (\i \DSlSl)\psi$\, or equivalently $\langle(\i\PtSl + g(W+T))\cdt \psi \opsi \rangle$\,, including a gauge coupling constant\footnote{In the Yang-Mills complex self-dual antisymmetric tensor model, the form $(\i\PtSl + gW-gT)$\,
is required, in order that the covariant curvatures be formed correctly (with a single coupling constant $g$, in contrast to the expected two coupling constants for the gauge group $SU(2)\!\times\! U(1)$ at the classical level).}.
In the internal- and Dirac algebra-decorated form, the ``super-potential" contribution $\Phi = W+T$ is overall odd-graded, if account is taken of both the internal (superalgebra, ${\mathbb Z}_2$ parity) and external (Clifford, $\gamma_5$-parity) grading, consistent with \cite{ThierryMieg2021newchi} (c.f. \cite{quillen1985superconnections}).  
These constructions, within the scope of the present approach, will be taken up in subsequent work.

In conclusion, we should note that significant impediments persist for the satisfactory formulation of local field theories  
with internal supersymmetry (and in particular, in the applications to model building, in our complex antisymmetric tensor model): firstly, the indefinite sign of the even part of the superalgebra Killing form implies non-positive kinetic energy for the gauge sector at the classical level, and secondly, scalar multiplets and symmetry breaking are not addressed in the present work. 
It should also be pointed out that the component field transformations 
(\ref{eq:DeltaCcliff}), (\ref{eq:DeltaCcpts}), (\ref{eq:DeltaCcliffComplex}) and 
(\ref{eq:DeltaCcliffComplex2}) induced by the additional odd-graded $SU(2/1)$ internal generators are not consistently implemented, on other-than-linear monomials:  
referring to equation (\ref{eq:FrstOrderLagAuxOnly}), the cancellation resulting from
\smash{$\delta_K\big(b^{I\rho}\big)\ob^{\oJ}{}_\rho \kappa_{I \oJ}$} is maintained under $\delta_{\oK}$\,, with the same ratio of mass scaling parameters $\ttm_B$ and $\ttm_{\oB}$ to $\ttm_W$, only if we assume that the correct\footnote{A more formal framework than that developed here, might necessitate the symmetries being realized consistently only on the complex dual $U(sl(2/1))^*$\,, rather than the polynomial algebra ${\mathbb C}[sl(2/1)^*]$.}  variation is \smash{$-b^{I\rho} \delta_{\oK}\big(\ob^{\oJ}{}_\rho\big) \kappa_{I \oJ}$}\,.

%\vfill
\noindent
\textbf{Acknowledgements:}\\
%The authors wish to thank \\
% This research was supported in part by the Intramural Research Program of the National Library of Medicine, National Institute of Health.
 This research was supported in part by the Intramural Research Program of the National Institutes of Health (NIH). The contributions of the NIH author are considered Works of the United States Government.
\bibliographystyle{unsrt}
\bibliography{YMACLit}

\mbox{}

\noindent
\begin{appendix}
\section{Derivations} 
\label{sec:Derivations}

\noindent
\textbf{Conventions:}\\
Coordinates in Minkowski space are labelled $\mu,\nu = 0,1,2,3$ with metric signature $\eta_{\mu\nu} = \mathrm{diag}(+1,-1,-1,-1)$\,. The Dirac algebra is generated by matrices $\gamma_\mu$ satisfying the Clifford algebra $\{\gamma_\mu,\gamma_\nu\}=2 \eta_{\mu\nu}$\, as usual. Lorentz generators are $\sigma_{\mu\nu} = \tfrac12\mathrm{i} [\gamma_\mu, \gamma_\nu]$\,.
The pseudoscalar $\gamma_5 := +\mathrm{i} \gamma^0\gamma^1\gamma^2  \gamma^3$\,, or
\smash{$\gamma_5 =+{\frac{\mathrm{i}}{4!}}\varepsilon^{\mu\nu\rho\sigma}\gamma_\mu \gamma_\nu \gamma_\rho \gamma_\sigma$\,}\,,
with  $\varepsilon_{0123} = +1$\, the completely antisymmetric Levi-Civita symbol.
We note the following rearrangement and trace identities
\begin{align}
\label{eq:SigGamId}
%[\gamma^\rho,[\gamma^\mu,\gamma^\nu]]=&\,4 (\eta^{\rho\mu}\gamma^\nu -  \eta^{\rho\nu}\gamma^\mu) \\
%\{\gamma^\rho,[\gamma^\mu,\gamma^\nu]\}=&\,+4\i\varepsilon^{\rho\mu\nu\alpha}\gamma_\alpha \gamma^5\,;\\
 [[\gamma^\mu,\gamma^\nu],\gamma^\rho]=&\,4(\eta^{\nu\rho}\gamma^\mu -  \eta^{\mu \rho}\gamma^\nu)\,,\\
 \{[\gamma^\mu,\gamma^\nu],\gamma^\rho\}=&\, -4\i\varepsilon^{\mu\nu\rho\alpha}\gamma_\alpha \gamma_5\,;
 \nonumber\\
\Tr \big(\sigma^{\mu\nu}\gamma^\rho\gamma^\sigma\big) = &\, 4\i(\eta^{\nu\rho}\gamma^\mu -  \eta^{\mu \rho}\gamma^\nu)\,,\\
\Tr \big( \sigma^{\mu\nu}\gamma^\rho\gamma^\sigma\gamma_5\big) = &\, 
=-4\varepsilon^{\mu\nu\rho\sigma}\,.\nonumber
 \end{align}
\noindent
%\mbox{}\\
%\vfill
%\pagebreak
\textbf{Lagrangian \boldmath{${\mathscr L}_{YMCT}$}}
\begin{description}  
\item[\boldmath{$ \partial W\omega         $}]\mbox{}
    \begin{align*}
    \tfrac 12 \partial^\mu W^{m\nu} \omega^{n \rho\sigma} \langle \gamma_\mu\gamma_\nu \sigma_{\rho\sigma}\rangle\langle \Lambda_m.\Lambda_n\rangle = &\, \tfrac 12 \partial^\mu W^{m\nu} \omega^{n \rho\sigma} 
    4\i(\eta_{\nu\rho}\eta_{\mu\sigma}- \eta_{\mu\rho}\eta_{\nu\sigma})\langle\Lambda_m.\Lambda_n\rangle \\
    =&\, -4\i\tfrac 12(\partial^\mu W^{m\nu}  -\partial^\nu W^{m\mu}) \omega^{n}_{\mu\nu} \langle\Lambda_m.\Lambda_n\rangle  
    \end{align*}
\item[\boldmath{$W W\omega         $}]\mbox{}
    \begin{align*}
    \tfrac 12 W^{k\mu} W^{m\nu} \omega^{n \rho\sigma} \langle \gamma_\mu\gamma_\nu \sigma_{\rho\sigma}\rangle\langle \Lambda_m.\Lambda_n\rangle = &\, \tfrac 12 W^{k\mu} W^{m\nu} \omega^{n \rho\sigma}
        4\i(\eta_{\nu\rho}\eta_{\mu\sigma}- \eta_{\mu\rho}\eta_{\nu\sigma})\langle\Lambda_k\Lambda_m.\Lambda_n\rangle \\
        =&\, -4\i (\tfrac 12 W^{k\mu} W^{m\nu} - \tfrac 12W^{k\nu} W^{m\mu})\omega^{n}_{\mu\nu}\langle\Lambda_k\Lambda_m.\Lambda_n\rangle\\
        &\, -4\i (\tfrac 12 W^{k\mu} W^{m\nu}\omega^{n}_{\mu\nu})\langle[\Lambda_k,\Lambda_m].\Lambda_n\rangle\,.
    \end{align*}   
    \item[\boldmath{$ \partial W\omega + W W\omega $}]\mbox{}  
    \begin{align*}
       \langle\partial W\omega + W W\omega \rangle =&\, 
       -4\i \tfrac12(\partial^\mu W^{m\nu} -\partial^\nu W^{m\mu}+ W^{k\ell}W^{\ell\nu}C_{k\ell}{}^m)\omega^n{}_{\mu\nu})\langle \Lambda_m.\Lambda_n\rangle
           \end{align*}  
  \item[\boldmath{$ \omega \cdt \omega $}]\mbox{}  
    \begin{align*}
  \omega^{m \kappa\lambda} \omega^{n \mu \nu}\tfrac14 \langle \sigma_{\kappa\lambda} \sigma_{ \mu \nu}\rangle
  \langle \Lambda_m.\Lambda_n\rangle
  = &\,\omega^{m \kappa\lambda} \omega^{n \mu \nu} \langle \Lambda_m.\Lambda_n\rangle
\tfrac14. 4(\eta_{\kappa\mu}\eta_{\lambda\nu} - \eta_{\kappa\nu}\eta_{\lambda\mu} ) =
2 \omega^{m \mu \nu} \omega^{n}_{\mu \nu} \langle \Lambda_m.\Lambda_n\rangle\,;\\  
\mbox{Finally}\qquad 
 {\mathscr L}_{YM} :=&\, \langle \tfrac 12\i(\PtSl + W)W \cdt \omega\rangle + \langle\omega\cdt\omega\rangle
 = \big(F^{m\mu\nu}(W)\omega^n_{\mu\nu} + 2 \omega^{m \mu \nu} \omega^{n}_{\mu \nu}\big)\eta_{mn}
   \end{align*} 
 (whence\footnote{Completing the square and removing a Gaussian from the generating functional, or eliminating 
fields using the algebraic constraints giving $\omega = -\tfrac 14 F$\,.} 
  %   \quad \omega^m_{\mu\nu}= &\, -\tfrac14F^{m\mu\nu}\,, \quad 
 ${\mathscr L}_{YM} =  -\tfrac 18 F^{m\mu\nu}F^{n}_{\mu\nu}\zeta_{mn}$\quad 
 after elimination)
 \end{description}  
% while
% whence (for ${\mathscr L}_{YMCT}$)
\begin{description}
\item[\boldmath{$ \PtSl T t+  (W T-TW)t $}]\mbox{}
\begin{align*}
\langle\PtSl T t\rangle =&\, \partial^\rho T^{a\mu\nu}. t^b{}_\sigma
\langle \gamma_\rho\sigma_{\mu\nu}\gamma_\sigma
\rangle \langle \Lambda_a \cdt \Lambda_b\rangle \\
=&\, \partial^\rho T^{a\mu\nu}. t^b{}_\sigma (4\i)(\eta_{\mu\rho}\eta_{\nu\sigma} - \eta_{\nu\rho}\eta_{\mu\sigma} ) \zeta_{ab} = 8\i \partial_\mu T^{a\mu\nu}.t^b{}_\nu \zeta_{ab}\,
\end{align*}
\mbox{}\\[-1cm]
\begin{align*}
\langle(W T-TW)t \rangle =&\, W^{m\rho} T^{a\mu\nu}. t^b{}_\sigma
\big(\langle \gamma_\rho\sigma_{\mu\nu}\gamma_\sigma \rangle 
\langle \Lambda_m \Lambda_a \cdt \Lambda_b\rangle
+  \langle \sigma_{\mu\nu} \gamma_\rho\gamma_\sigma \rangle 
\langle \Lambda_a \Lambda_m \cdt \Lambda_b\rangle\\
=&\, W^{m\rho} T^{a\mu\nu}. t^b{}_\sigma \big(
(4\i)(\eta_{\mu\rho}\eta_{\nu\sigma} - \eta_{\nu\rho}\eta_{\mu\sigma} )\langle \Lambda_m \Lambda_a \cdt \Lambda_b \rangle \\
&\,+(4\i)(\eta_{\nu\rho}\eta_{\mu\sigma} - \eta_{\mu\rho}\eta_{\nu\sigma} )(\langle \Lambda_a \Lambda_m \cdt \Lambda_b \rangle \big)\\=&\, 
8\i W^{m}_\mu T^{a\mu\nu}.t^b{}_\nu\langle \Lambda_m \Lambda_a \cdt \Lambda_b \rangle
+8\i W^{m}_\nu T^{a\mu\nu}.t^b{}_\mu\langle \Lambda_a \Lambda_m \cdt \Lambda_b \rangle\\
=&\, 8\i W^{m}_\mu T^{a\mu\nu}.t^b{}_\nu \langle [\Lambda_m, \Lambda_a] \cdt \Lambda_b \rangle
\equiv 8\i W^{m}_\mu T^{a\mu\nu}.t^b{}_\nu C_{ma}{}^c \zeta_{cb}
\end{align*}

Now define $(D_\mu T)^{a\mu\nu} = \partial_\mu T^{a\mu\nu} + W^{m}{}_\mu T^{c\mu\nu} C_{mc}{}^a$, giving
\begin{align*}
\langle(\PtSl T+[W,T])t\rangle=&\, 8\i (D_\mu T)^{a\mu\nu} t^b{}_\nu \kappa_{ab}\,;\\
%\end{align*}
%%%so from ${\mathscr L}_{YM}$ noting
%%also
% \begin{align*}
 \langle t \cdt t\rangle = &\,t^{a\mu}t^{b \nu}\langle \gamma_\mu\gamma_\nu \rangle \langle\Lambda_a\cdt\Lambda_b\rangle = 4 t^{a\mu}t^{b}{}_{ \nu}\zeta_{ab}\,.
 \end{align*}
We have
\footnote{Taking $\lambda\langle (\PtSl+W-T)\cdt T\rangle + \langle t\cdt t\rangle$
gives $8\i\lambda DT\cdt t+ 4t\cdt t$\,, leading to ${\mathscr L} = 4\lambda^2 (DT)^2$.}
\begin{align*}
{\mathscr L}_{CT} := &\,\tfrac12\i \langle (\PtSl T + \{W,T\})\cdt t\rangle + \langle t \cdt t \rangle\\
= &\,-4 (D_\mu T)^{a\mu\nu} t^b{}_\nu \kappa_{ab}+ 4 t^{a\mu}t^{b}{}_{ \nu}\zeta_{ab}
\end{align*}
(whence  ${\mathscr L}_{CT} =\tfrac 12 (D_\rho T)^{a\rho \nu} (D_\sigma T)^{b\sigma}{}_{ \nu} \zeta_{ab}$
after elimination )\,.%$k=\tfrac 14 DT$
 \end{description}   
 \mbox{}
 
 \noindent
 \textbf{Off shell invariance:}\\
 In section \ref{sec:Superalgebra} above we presented a set of algebraic-level component field transformation 
 rules based on the closure of the ``Clifford field'' expansion  (\ref{eq:CliffSplits}) under the action
 (via either commutator or graded commutator brackets) of the internal generators $\Lambda_c$ (or
 $\Lambda_I, \oLambda_{\oI}$)\,. The invariance of the second order form of the Lagrangian ${\mathscr L}_{YMCT}$
 (equation (\ref{eq:finalLymacSecond}), (\ref{eq:YMCTsuper}))  was demonstrated under such field transformations with the help of the dynamical equations of motion. Here we present 
the equivalent demonstration for the 
 action (\ref{eq:finalL_ymac_first}), (\ref{eq:superL1complex}) in the first order setup. 
 Rather than use the algebraic constraints to
 eliminate the auxiliary component fields $\omega$ and $t$ in terms of curvatures, we here reverse the procedure and take as a starting point, the action expressed only in terms of these auxiliaries.
 
For the $SU(3)$ version we have from (\ref{eq:finalLymacSecond}), 
\[
{\mathscr L}_{YMCT}^1 = \texttt{f}\,\omega^{m \rho \sigma}\omega^{n}{}_ {\rho \sigma}\eta_{mn} + 
\texttt{g} \, t^{a\rho} t^{b}{}_\rho \eta_{ab}\,,
\]
and introducing, for clarity, running constant coefficients $\texttt{f}=-2$\,, $\texttt{g}=-4$\, coming from the elimination of auxiliary fields in the Lagrangian (which gave $\omega := \texttt{w} F =-\tfrac 14F$\,,
$t :=\btau D\cdt T = +\tfrac 12 D\cdt T$\,). At this point it is necessary to introduce explicit formal mass scaling parameters 
$W\rightarrow \ttm_{W} W$\,, $T\rightarrow \ttm_{T} T$  into the Clifford expansions 
(\ref{eq:CliffSplitsNo3}) of $\Phi_\gamma$ and 
 $\Phi_\sigma$\,, in order for the physical dimensions to match under field reshufflings. Using the mass parameter re-scaled equivalents of the component variations (\ref{eq:DeltaCcpts}), and integration by parts, we find
\begin{align*}
\delta_c\big({\mathscr L}^1_{YMCT}\big) = 
%&\,2 \texttt{f} (\ttm_T C_{ca}{}^m T^{a \rho \sigma}) \omega^n{}_{\rho \sigma}\eta_{mn}
%+ 2 \texttt{g}(\ttm_W  \tau C_{cm}{}^a W^{m \rho}) t^b{}_\rho \eta_{ab}\\
=&\, 2 \texttt{f} (\ttm_T C_{ca}{}^m T^{a \rho \sigma}) \omega^n{}_{\rho \sigma}\eta_{mn}
+ 2 \texttt{g}(\ttm_W  C_{cm}{}^a W^{m \rho}) t^b{}_\rho \eta_{ab}\\
=&\, 2 \texttt{f} \ttm_T (C_{cam} T^{a \rho \sigma}) (\ttw F^m{}_{\rho \sigma})
+ 2\texttt{g} \ttm_W  (C_{cma} W^{m \rho})(\btau  D\cdt T)^a{}_\rho \\
=&\, (-4 \texttt{f} \ttw \ttm_T C_{cam}  + 2\texttt{g} \btau \ttm_W  C_{cma})(D\cdt T)^{a \rho}W^{m}{}_{\rho}\,.
\end{align*}
which (as $C_{cma}=-C_{cam}$) vanishes if the mass scaling parameters satisfy
$\ttm_T = 2 \ttm_W$\,.

The $SU(2/1)$ case proceeds analogously, starting with (\ref{eq:YMCTsuper}). From the algebraic elimination
we have again $\omega := \texttt{w}F =-\tfrac 14F$\,, $b :=\btau D\cdt B \equiv  +\tfrac 12 D\cdt B$\,, but in complex coordinates $\ob := \overline{\btau} D\cdt \oB \equiv  -\tfrac 12 D\cdt \oB$\,, 
in view of the distinction between the $\eta$ and $\kappa$ bilinear forms
(appendix \ref{app:NotationSU21SU3})\,. We have
\begin{align}
\label{eq:FrstOrderLagAuxOnly}
{\mathscr L}_{YMCT}^1 = \texttt{f}\,\omega^{m\rho\sigma}\omega^n{}_{\rho\sigma}\kappa_{mn} + 
2\texttt{g} b^{I\rho} \ob^{\oJ}{}_\rho \kappa_{I \oJ}\,.
\end{align}
with $\texttt{f}=-2$\,, $\texttt{g} = -4 $\,. Evaluating \smash{$2\texttt{f}\delta_K(\omega^{m\rho\sigma})\omega^n{}_{\rho\sigma}\kappa_{mn} \!+\!
2\texttt{g} \delta_K\big(b^{I\rho}\big)\ob^{\oJ}{}_\rho \kappa_{I \oJ}$}\, leads to
 \begin{align*}
\delta_K\big({\mathscr L}^1_{YMCT}\big) = &\, 
2 \texttt{f}\,(C_{K\oJ}{}^{m} \texttt{m}_{\oB} {\oB}^{\oJ \rho\sigma} )\omega^n_{\rho\sigma}\eta_{mn} + 2\texttt{g} \, (C_{Km}{}^I {\texttt{m}_W}W^{m\rho}) \ob^{\oJ}{}_\rho \kappa_{I\oJ}\\
 = &\, -4 \texttt{f} \texttt{w} \,(C_{K\oJ}{}^{m} \texttt{m}_{\oB} (D \cdt \oB)^{\oJ\rho} )W^n{}_{\rho}\kappa_{mn}  + 2\texttt{g} \overline{\btau}  \, (C_{Km}{}^I {\texttt{m}_W}
 W^{m\rho}) (D\cdt \oB)^{\oJ}{}_\rho \kappa_{I\oJ}\\
 =&\, (-4 \texttt{f} \texttt{w} \texttt{m}_{\oB} C_{K\oJ m}+2\texttt{g} \overline{\btau}  {\texttt{m}_W}C_{Km\oJ})(D \cdt \oB)^{\oJ\rho} W^m{}_{\rho}\,,
 %=&\, ((-4)(-2)(-\tfrac 14)\ttm_{\oB} -2(-4)(-\tfrac12) \ttm_W)C_{K\oJ m}(D \cdt \oB)^{\oJ\rho} W^m{}_{\rho}
\end{align*}
using $C_{Km\oJ}=-C_{K\oJ m}$\,, with cancellation when $\ttm_{\oB}= -2 \ttm_W$\,. Analogously, evaluation of
\smash{$2\texttt{f}\delta_{\oK}(\omega^{m\rho\sigma})\omega^n{}_{\rho\sigma}\kappa_{mn} \!+\!
2\texttt{g} b^{I\rho} \delta_{\oK}\big(\ob^{\oJ}{}_\rho\big) \kappa_{I \oJ}$}, using $C_{Km J}=-C_{K J m}$, leads to cancellation when
$\ttm_{B}= +2 \ttm_W$\,.
%\mbox{}\\

%\pagebreak
\section{$SU(3)$ and $SU(2/1)$ generators and structure constants.} 
\label{app:NotationSU21SU3} 

\noindent
In the main text we have developed a portmanteau notation for the model symmetries, both for the gauge group
$G \!\times\!U(1)$ (for the special choice $G=SU(2)$)\,, and for the generators and field transformations associated with a possible underlying higher symmetry, with index enumeration $\Lambda_m$\, and $\Lambda_a$ (or $\Lambda_I\,,\oLambda_{\oJ}$)
associated with the gauge group and additional transformations, respectively, leading to stucture constants
$C_{mn}{}^p$\,, $C_{ma}{}^b$\,, $C_{ab}{}^m\,,\cdots$,
using identical notation for each candidate higher symmetry in question ($SU(3)$ versus $SU(2/1)$). Here we provide details specific to differentiating between each of these cases, regarding the nomenclature, structure relations and invariant forms.

\noindent
\textbf{Lie algebra \boldmath{$SU(3)$}:}\\
The $3 \!\times\!3$ Gell-Mann matrices provide an hermitean basis for generators of either the Lie algebra $SU(3)$, (or as we shall see, with a slight modification, the Lie superalgebra $SU(2/1)$)\,. 
%For $SU(3)$ the set $\{\lambda_{\sf I}, \, {\sf I} =1,2,\cdots, 8\}$ comprises the 
For $SU(3)$ set $\Lambda_m\,, m= 1,2,3,8$ provide the basis of generators of the $SU(2)\times U(1)$ gauge algebra\,, namely  $\tfrac12\lambda_i\,, i=1,2,3$\, together with $\tfrac12\lambda_8$,
\begin{align}
\label{eq:TripletLambdas} 
\lambda_1  = &\, \scalebox{.9}{$\left(\begin{array}{ccc} 0&1&0\\1&0&0\\0&0&0\end{array}\right)$},\,
\lambda_2  = \scalebox{.9}{$\left(\begin{array}{ccc} 0&-\i&0\\ \i &0&0\\0&0&0\end{array}\right)$},\,
\lambda_3  =\scalebox{.9}{$\left(\begin{array}{ccc} 1&0 &0\\0 & -1 &0\\0&0&0\end{array}\right)$},\,
\lambda_8  =\scalebox{.9}{$\frac{1}{\sqrt{3}}\left(\begin{array}{ccc} 1&0&0\\0&1&0\\0&0&-2\end{array}\right)$}.
\end{align}
together with  off-diagonal generators (Gell-Mann matrices $\tfrac 12 \lambda_a\,, a=4,5,6,7$) given in terms of the complex basis as $ \Lambda_I\,, \oLambda_{\oI}\,,I=\emO,\emT\,, \oI = \oemO,\oemT$\, defined by\footnote{For the sake of uniformity 
we take $ \Lambda_I\equiv \lambda_I\,, \oLambda_{\oI}\equiv\olambda_{\oI}$ and we regard the
$\Lambda$ as providing the abstract definition.}:
\begin{align*}
\lambda_{\emO} := &\, \tfrac12\lambda_{4}+ \tfrac12\i \lambda_{5}\,,  \quad 
\lambda_{\emT} := \tfrac12\lambda_{6}+ \tfrac12\i \lambda_{7}\,,\\
\olambda_{\oemO} := &\, -\tfrac12\lambda_{6}+\tfrac 12 \i \lambda_7=-(\lambda_{\emT})^\dagger\,, \quad
 \olambda_{\oemT} := +\tfrac12\lambda_{4}-\tfrac 12 \i \lambda_5 = +(\lambda_{\emO})^\dagger\,,
\end{align*}
\begin{align}
\label{eq:OddComplexLambdas}
\lambda_{\emO}=&\, \scalebox{.9}{$\left(\begin{array}{ccc} 0&0&1\\0&0&0\\0&0&0\end{array}\right)$}, \,
\lambda_{\emT} = \scalebox{.9}{$\left(\begin{array}{ccc} 0&0&0\\ 0 &0&1\\ 0&0&0\end{array}\right)$},\, 
\olambda_{\oemO}  = \scalebox{.9}{$\left(\begin{array}{ccc} 0&0 &0\\0 &0 &0\\ 0 &-1&0\end{array}\right)$},\,
\olambda_{\oemT} = \scalebox{.9}{$\left(\begin{array}{ccc} 0&0 &0\\0 &0 &0 \\1 &0 &0 \end{array}\right)$},\,
\end{align}
defined such that both pairs transform like standard doublets of $SU(2)$\,. A uniform index notation for the
basis is $\{\Lambda_{\sf I}, \, {\sf I} =1,2,\cdots, 8\}$\,.

\noindent
\textbf{$SU(3)$ Killing form:}\\
The $SU(3)$ Killing trace form $\eta_{\sf{IJ}}= \Tr(\lambda_{\sf I} \lambda_{\sf J})$ has nonzero components
\begin{align}
\label{eq:KillingEvenLA}
\eta_{ij} =&\, \Tr(\lambda_i\lambda_j)=2\delta_{ij}\,, \quad 
\eta_{88} = \Tr(\lambda_8\lambda_8)=2\,
%\eta_{ab} =&\, \Tr(\lambda_a\lambda_b)=2\delta_{ab}\,, \quad 
\end{align}
and $\eta_{ab} = \Tr(\lambda_a\lambda_b)=2\delta_{ab}$\,,
or in the complex basis $ \lambda_I\,, \olambda_{\oI}\,,I=\emO,\emT\,, \oI = \oemO,\oemT$\,
$\eta_{I\oJ} = \Tr(\lambda_I \olambda_{\oJ})=\eta_{\oJ I}$\,,
\begin{align}
\label{eq:KillingOddLAComp}
\Tr(\lambda_{\emO} \olambda_{\oemT} ) = &\,+1\,,\quad
\Tr(\lambda_{\emT}  \olambda_{\emO} ) = -1\,;
\quad 
\Tr( \olambda_{\oemO} \lambda_{\emT}) = -1\,,\quad
\Tr(\olambda_{\oemT}\lambda_{\emO}  ) = +1\,.
\end{align}
These quantities can also be written 
$ \eta_{I \oJ} = \varepsilon_{I\oJ}\,, 
\eta_{\oI J} = \varepsilon_{J\oI}\,$,
with the $2\!\times\!2$ Levi-Civita symbol being
$\varepsilon_{\emO\oemT} = 1 = - \varepsilon_{\oemT\emO}$\,,
$\varepsilon_{\emO\oemO} = \varepsilon_{\oemO\emO}$ $=
\varepsilon_{\emT\oemT} = \varepsilon_{\oemT\emT} =0$\,.
The Killing form is overall symmetric, with Euclidean signature and block diagonal form, the second $4\times 4$ 
block in the complex basis being an off-diagonal matrix:
\begin{align}
\label{eq:EtaRealFbyF}
\eta_{ab}\big|_{a,b=4}^7 = \scalebox{.9}{$\left(\begin{array}{cccc} 2&0&0&0\\ 0&2&0&0\\
0&0&2&0\\ 0&0&0&2\end{array}\right)$}\,, \quad \mbox{or}\quad
\left.\left(\begin{array}{cc} \eta_{ I J} &\eta_{I\oJ}\\
\eta_{\oI J} & \eta_{\oI\oJ}\end{array}\right)\right|_{\emO,\oemO}^{\emT, \oemT} = 
\scalebox{.9}{$\left(\begin{array}{cccc} 0&0&0&1\\ 0&0&-1&0\\
0&-1& 0&0\\ 1&0&0&0\end{array}\right)$}\,.
\end{align}

\noindent
\textbf{Lie superalgebra $SU(2/1)$:}\\
For $SU(2/1)$ the set $\Lambda_m\,, m= 0,1,2,3$ provides the basis of generators of the $SU(2)\times U(1)$ gauge algebra\,, namely  $\tfrac12\lambda_i\,, i=1,2,3$\, as above, together with the super-traceless matrix $\lambda_0 := \mathrm{diag}(1,1,2)$\,.
As for $SU(3)$ above we set $\Lambda_a := \tfrac 12\lambda_a$\, or in complex 
combinations\footnote{ Together with the $\lambda_i$ and $\lambda_a$\,, any two of the diagonal matrices
$\lambda_8\,, \lambda_0$ and the super-identity $\chi =\mathrm{diag}(1,1,-1)$ provide a basis for the associative algebra of $3\!\times\!3$ matrices (equations
(\ref{eq:TripletLambdas}), (\ref{eq:OddComplexLambdas}))\,.} \smash{$\Lambda_I\equiv \lambda_I\,, \oLambda_{\oI}\equiv \olambda_{\oI}$}. A uniform index notation for the
basis is $\{\Lambda_{A}, \, A =0,1,2,\cdots, 7\}$\,.

The Lie superalgebra
is defined by its closure (commutation and anticommutation) relations under the graded bracket, 
$[\Lambda_A,\Lambda_B] := \Lambda_A\Lambda_B -[AB] \Lambda_B\Lambda_A$\, in concrete representations, where the grading factors 
are defined $[A] = (-1)^{|A|}$\,, $[AB] = (-1)^{|A||B|}$, with $|A| = 0$ for even generators, and $|A|=1$ for odd generators (that is, $[m] = +1$, $m = 0,1,2,3$ and $[a]= -1$, $a = 4,5,6,7 $ (or $[I] =[\oJ] = -1\,, I=\emO,\emT, \oJ = \oemO,\oemT$). 

\noindent
\textbf{$SU(2/1)$ Killing form:}\\
The $SU(2/1)$ supertrace Killing form $\kappa_{AB}= \Str(\lambda_A \lambda_B)$ has even-even and odd-odd components 
\begin{align}
\label{eq:KillingEven}
\kappa_{ij} = \Str(\lambda_i\lambda_j)=2\delta_{ij}\,, \quad &\,
\kappa_{00} = \Str(\lambda_0\lambda_0)=-2\,, \\
\Str(\lambda_{\emO} \olambda_{\oemT} ) = +1\,,\quad
\Str(\lambda_{\emT}  \olambda_{\emO} ) = -1\,;
\quad &\,
\Str( \olambda_{\oemO} \lambda_{\emT}) = +1\,,\quad
 \Str(\olambda_{\oemT}\lambda_{\emO}  ) = -1\,, \nonumber
\end{align}
respectively. These quantities can also be written 
$ \kappa_{I \oJ} = \varepsilon_{I\oJ}\,, 
\kappa_{\oI J} = -\varepsilon_{J\oI}\,$, in terms of the Levi-Civita symbol as above.
Note that the Killing form is even and orthosymplectic:  the $4\times 4$ even-even part has Minkowski form $2\,\mathrm{diag}(-1,1,1,1)$\,, while (in either real or complex basis), the odd-odd part  
is a  $4\times 4$ antisymmetric matrix (compare (\ref{eq:EtaRealFbyF}) )
\begin{align}
\label{eq:KappaRealFbyF}
\kappa_{ab}\big|_{a,b=4}^7 = \scalebox{.9}{$\left(\begin{array}{cccc} 0&2\i&0&0\\ -2\i&0&0&0\\
0&0&0&2\i\\ 0&0&-2\i &0\end{array}\right)$}\,, \quad \mbox{or}\quad
\left.\left(\begin{array}{cc} \kappa_{ I J} &\kappa_{I\oJ}\\
\kappa_{\oI J} & \kappa_{\oI\oJ}\end{array}\right)\right|_{\emO,\oemO}^{\emT, \oemT} = 
\scalebox{.9}{$\left(\begin{array}{cccc} 0&0&0&1\\ 0&0&-1&0\\
0&1&0&0\\ -1&0&0&0\end{array}\right)$}\,. 
\end{align}

\noindent
\textbf{Structure constants:}\\
For the $SU(3)$ Lie algebra the commmutation relations of the generators $\Lambda_I=\tfrac 12 \lambda_{\sf I}$
are given by the standard Gell-Mann $f$ symbols, $[\tfrac12\lambda_{\sf I}, \tfrac12\lambda_{\sf J}]= 
\sum_{\sf K} \mathrm{i}f_{{\sf I}  {\sf J}{\sf K}} \tfrac12\lambda_{\sf I} $\,. 
In the text we adopt the definition
\smash{$[\Lambda_{\sf I}, \Lambda_{\sf J}]=\left.\sum\right._K C_{{\sf I}{\sf J}}{}^{\sf K} \Lambda_{\sf K}
$}. 
In particular we have $C_{ij}{}^k := \mathrm{i}f_{ijk}$\,, and   \smash{$C_{i8}{}^k= C_{8j}{}^k=0$}\,.
For $SU(2/1)$ the graded commutation relations are defined as \smash{$[\Lambda_A,\Lambda_B]=\left.\sum\right._C C_{AB}{}^C\Lambda_C$}\,. Once again we have $C_{ij}{}^k = \mathrm{i}f_{ijk}$ as in the $SU(2)$ case. In appendix \ref{sec:SprTensor} below
we provide brief information of graded tensor notation for superalgebras, the properties of the Killing form
and the covariant graded structure tensor $C_{ABC}$\,. 

The following table provides nonzero structure constants $C_{{\sf I}{\sf J}}{}^{\sf K}$\, and $C_{AB}{}^C$\,, namely
\smash{$C_{mI}{}^J$\,, $C_{m\oI}{}^{\oJ}$} and \smash{$C_{I {\oJ}}{}^m =\mp C_{{\oJ}I}{}^m$}, with 
the algebra and superalgebra distinguished on entries $_8$ or $_0$\,:
%$SU(3)$:
\begin{align*}
SU(3): \,&\,
\begin{tabular}{c||cccccc}
\hline
$C_{8I}{}^J$ & $C_{3{\emO}}{}^{{\emO}}$ & $C_{1{\emO}}{}^{{\emT}}$ & $C_{1{\emT}}{}^{{\emO}}$ & 
$C_{3{\emT}}{}^{{\emT}}$ & $C_{2{\emT}}{}^{{\emO}}$ & $C_{2{\emO}}{}^{{\emT}}$\\%%%%%
 %& $C_{1{\emT}}{}^{{\emT}}$ & $C_{2{\emO}}{}^{{\emT}}$ & $C_{2{\emO}}{}^{{\emT}}$\\
$\scalebox{.9}{$\frac{\sqrt{3}}{2}$}\,\delta_I{}^J$ &
$+\tfrac 12$&$+\tfrac 12$&$+\tfrac 12$& $-\tfrac 12$ &$-\tfrac 12\i$& $+\tfrac 12\i $\\[.2cm]
\hline 
$C_{8{{\oI}}}{}^{{\oJ}}$ & $C_{3{\oemO}}{}^{{\oemO}}$ & $C_{1{\oemO}}{}^{{\oemT}}$ & $C_{1{\oemT}}{}^{{\oemO}}$ & 
$C_{3{\oemT}}{}^{{\oemT}}$ & $C_{2{\oemT}}{}^{{\oemO}}$ & $C_{2{\oemO}}{}^{{\oemT}}$\\
$\scalebox{.9}{$-\frac{\sqrt{3}}{2}$}\,\delta_{\oI}{}^{\oJ}$ &
$+\tfrac 12$&$+\tfrac 12$&$+\tfrac 12$& $-\tfrac 12$ &$-\tfrac 12\i$& $+\tfrac 12\i $\\[.2cm]
\hline 
\end{tabular}\, \quad \begin{tabular}{cccc}
\hline
$C_{{\emO}{\oemO}}{}^{1}$ & $C_{{\emO}{\oemO}}{}^2$ & $C_{{\emO}{\oemT}}{}^3$ & $C_{{\emO}{\oemT}}{}^8$ \\
$-1\!\!$ & $-\i\!\!$ &$1\!\!$ &$\sqrt{3}\!\!$\\
\hline
 $C_{{\emT}{\oemO}}{}^3$ & $C_{{\emT}{\oemO}}{}^8$ & $C_{{\emT}{\oemT}}{}^1$ & $C_{{\emT}{\oemT}}{}^2$ \\
$1\!\!$ &$-\sqrt{3}\!\!$& $1\!\!$& $\!\!-\i$\\
\hline
\end{tabular}
\end{align*} 
\\[.2cm]
%%%%%%%%%%%%
%%%%%%%%%%%%%%%%%%%%%SU2/1
%SU(2/1): \qquad &\,
\begin{align*}
SU(2/1): \,&\,
\begin{tabular}{c||cccccc}
\hline
$C_{0I}{}^J$ \,\,& $C_{3{\emO}}{}^{{\emO}}$ & $C_{1{\emO}}{}^{{\emT}}$ & $C_{1{\emT}}{}^{{\emO}}$ & 
$C_{3{\emT}}{}^{{\emT}}$ & $C_{2{\emT}}{}^{{\emO}}$ & $C_{2{\emO}}{}^{{\emT}}$\\
$-\delta_I{}^J$ &
$+\tfrac 12$&$+\tfrac 12$&$+\tfrac 12$& $-\tfrac 12$ &$-\tfrac 12\i$& $+\tfrac 12\i $\\[.2cm]
\hline 
$C_{0{\oI}}{}^{\oJ}$ \,\,& $C_{3{\oemO}}{}^{{\oemO}}$ & $C_{1{\oemO}}{}^{{\oemT}}$ & $C_{1{\oemO}}{}^{{\oemT}}$ & 
$C_{3{\oemT}}{}^{{\oemT}}$ & $C_{2{\oemT}}{}^{{\oemO}}$ & $C_{2{\oemO}}{}^{{\oemT}}$\\
$+\delta_{\oI}{}^{\oJ}$ &
$+\tfrac 12$&$+\tfrac 12$&$+\tfrac 12$& $-\tfrac 12$ &$-\tfrac 12\i$& $+\tfrac 12\i $\\[.2cm]
\hline 
\end{tabular}\, \quad \begin{tabular}{cccc}
\hline
$C_{{\emO}{\oemO}}{}^{1}$ & $C_{{\emO}{\oemO}}{}^2$ & $C_{{\emO}{\oemT}}{}^3$ & $C_{{\emO}{\oemT}}{}^0$ \\
$-1\!\!$ & $-\i\!\!$ &$1\!\!$ &$\tfrac12\!\!$\\
\hline
$C_{{\emT}{\oemO}}{}^3$ & $C_{{\emT}{\oemO}}{}^0$ & $C_{{\emT}{\oemT}}{}^1$ & $C_{{\emT}{\oemT}}{}^2$ \\
$1\!\!$ &$-\tfrac12\!\!$& $1\!\!$& $\!\!-\i$\\
\hline
\end{tabular}
\end{align*}  
%%%%%%%%%
\textbf{Complex basis and bilinear forms:}\\
We give here explicit component field expansions over generators of the model internal symmetry
in the complex basis, which illustrate important properties of and differences between the Lie algebra and superalgebra cases. As illustrative example we consider the tensor Clifford field (c.f.  
(\ref{eq:FullCliff}), (\ref{eq:CliffSplitsNo3})) with components $\omega, T$ and suppress the 
accompanying Dirac $\sigma$ matrices. We take
\[
\Phi = \left.\sum\right._m \omega^m \Lambda_m + \left.\sum\right._{I} B^I \Lambda_I + 
\left.\sum\right._{\oI}\oB^{\oI} \oLambda_{\oI}
\]
with coordinates such that $\left.\sum\right._{I} B^I \Lambda_I + \left.\sum\right._{\oI}\oB^{\oI} \oLambda_{\oI}= 
\left.\sum\right._{a=4}^7 T^a \Lambda_a$\,, namely
\begin{align}
\label{eq:Bdefs}
B^{\emO} := &\,\tfrac12(T^{4}\!-\!\i T^5)\,, \,\,B^{\emT}:= \tfrac12(T^{6}\!-\!\i T^{7})\,, \,\,
%\oB^{\ottO} :=  -\tfrac12T^{6\!+\!\i 7}\,=-(B^\tttT)^\dagger\,,
%\oB^{\ottT} := +\tfrac12T^{4\!+\!\i 5} = +(B^\tttO)^\dagger\,,
\oB^{\oemO} :=-(B^\emT)^\dagger\,,
\oB^{\oemT} :=+(B^\emO)^\dagger\,.
\end{align}
%such that $\left.\sum\right._{I} B^I \Lambda_I + \left.\sum\right._{\oI}\oB^{\oI} \oLambda_{\oI}= 
%\left.\sum\right._{a=4}^7 T^a \Lambda_a$\,. 

We record various bilinear forms 
under the algebra or superalgebra, enumerating $\Phi$ above either as  $\Phi =\left.
\sum\right._{\sf I} \Phi^{\sf I}\Lambda_{\sf I}$ or $\Phi =\left.
\sum\right._{A} \Phi^{A}\Lambda_{A}$ in the  $SU(3)$ or $SU(2/1)$ coordinates, respectively.
Note that $\eta_{mn}=2\delta_{mn}$ for $m,n = 1,2,3,8$ in $SU(3)$ whereas $\kappa_{mn}$ has Minkowski signature
$2\mathrm{diag}(-1,1,1,1)$ for $m,n = 0,1,2,3$\, in $SU(2/1)$\,.
For the odd indices we use the abbreviation 
\smash{$
\varepsilon B \oB := \left.\sum\right._{I, \oJ} \varepsilon_{I \oJ}B^{I}\oB^{\oJ}
$}
(which equals \smash{$B^{\emO} B^{\emO}{}^\dagger + B^{\emT} B^{\emT}{}^\dagger$}). 
The following results are valid for local field combinations up to total derivatives:
\begin{align}
\label{eq:BilinearMagic}
 \left.\sum\right._{\sf I, \sf J} \Phi^{\sf I}\eta_{\sf I \sf J}  \Phi^{\sf J} =&\, 
\left.\sum\right.\omega^m\omega^n \eta_{mn} + 2\varepsilon B \oB\,, \quad  \left.\sum\right._{\sf I, \sf J} \Phi^{\sf I}\eta_{\sf I \sf J} \partial \Phi^{\sf J} =0 \,,\nn\\
 \left.\sum\right._{A,B} \Phi^{A}\kappa_{AB}  \Phi^{B} =&\,\left.\sum\right.\omega^m\omega^n \kappa_{mn} \,, \quad
\left.\sum\right._{A,B} \Phi^{A}\kappa_{AB}  \partial \Phi^{B} =2 \varepsilon B \partial \oB\,. 
\end{align}
\section{Tensor notation for superalgebras} 
\label{sec:SprTensor}
%\noindent 
For a Lie superalgebra $L=L_0 + L_1$ the ${\mathbb Z}_2$ graded bracket $[\cdot, \cdot]$ is antisymmetric or symmetric
(corresponding to a commutator or anticommutator in concrete matrix representations) according to the $\pm 1$ grading of homogeneous parts, namely $[X,Y]=-(-1)^{|X||Y|}[Y,X]\,,$ where $|[X,Y]| = |X| + |Y| \,(\mbox{mod 2})$, where the grading
homomorphism assigns $|X|=0,1$ for $X\in L_0$ or $X\in L_1$, respectively. The Jacobi identity asserts the
Leibniz property of the graded bracket\,,
\begin{align*}
 [X,[Y,Z]] =&\, [[X,Y],Z] + (-1)^{|X||Y|} [Y,[X,Z]]\,,
\end{align*}
and the adjoint action $\mathrm{ad}_X(Y):=[X,Y]$ provides a linear representation on the graded space $L$.
Below we use the abbreviations  $[X] = (-1)^{|X|}$\,, $[XY] = (-1)^{|X||Y|}$ for the 
${\mathbb Z}_2$ sign calculus.
For a fixed basis $\{ \Lambda_A \}$ we have the graded commutation relations
${[}\Lambda_A,\Lambda_B{]}= \left.\sum\right._C C_{AB}{}^C \Lambda_C\,$ and denoting the adjoint
of element $\Lambda_A$ as $\mbox{ad}_A$, we have representative matrices 
$\mbox{ad}_A (\Lambda_B):=\left.\sum\right._C  \Lambda_C (\mbox{ad}_A)^C{}_B$
given by $(\mbox{ad}_A)^C{}_B = -C_{AB}{}^C$\,. 

The bilinear Killing form is defined as the supertrace of the product of adjoint representation matrices\footnote{
For the $3\!\times\!3$ defining representation of  $SU(2/1)$ the supertrace form coincides with the Killing form up to proportionality (and similarly for the $SU(3)$ trace form)\,.} and we have
\begin{align*}
\kappa_{AB} := &\,\Str(\mbox{ad}_A \mbox{ad}_B)\, \equiv C_{AE}{}^F C_{BF}{}^E[F];\quad \\
\Str\big([\mbox{ad}_A \mbox{ad}_B, \mbox{ad}_C] \big)=&\, 0 = 
\Str\big(\mbox{ad}_A, [\mbox{ad}_B, \mbox{ad}_C ]\big)+
[BC]\Str\big([\mbox{ad}_A ,\mbox{ad}_C],  \mbox{ad}_B]\big)\\
\Leftrightarrow 0=&\, C_{BC}{}^E \kappa_{AE} + [BC]C_{AC}{}^E \kappa_{EB} 
\end{align*}
expressing its invariance (if it exists). Note that in the summation, $|A|=|E| \!+\! |F| =|B| \,(\mbox{mod 2})$\,, so
$\kappa$ is strictly diagonal, with even-even symmetric and odd-odd antisymmetric parts (graded symmetric, or ``orthosymplectic")\, 
so that $\kappa_{AB}= [AB] \kappa_{BA}=[A]\kappa_{BA} = [B]\kappa_{BA} $\,.
Thus $\kappa_{AB}$ is a graded totally symmetric 2-way tensor (of even type), while from above, the structure constants in covariant form
\[
C_{ABC}:= C_{AB}{}^E \kappa_{EC}
\]
constitute a graded  totally antisymmetric 3-way tensor\footnote{See \cite{dondi1981diagram}.}, satisfying
\[
C_{ABC}=-[AB]C_{BAC}=-[BC]C_{ACB} = +[AC][BC]C_{CAB}\,.
\] 
\mbox{}
\vfill

\end{appendix}
\end{document}